%% file: main.tex
\documentclass[sigconf,natbib=true]{acmart}

\usepackage{enumitem}
\usepackage{balance}
\usepackage{url}
\usepackage{amsmath}
\usepackage{mathtools,color}
\usepackage{graphicx}
\usepackage{multirow}
\usepackage{graphics}
\usepackage{algorithm}
\usepackage{algorithmic}

\usepackage{bm}
\usepackage[utf8]{inputenc}

\newcommand{\s}{\mathcal{S}}
\newcommand{\X}{\bm{X}}
\newcommand{\x}{\bm{x}}
\newcommand{\y}{\bm{y}}

\newcommand{\p}{{\bf{P}}}
\newcommand{\R}{\mathbb{R}}
\newcommand{\pe}{\bm{P}}
\newcommand{\rr}{\mathcal{R}}
\newcommand{\loss}{\mathcal{L}}
\newcommand{\U}{\bm{U}}
\usepackage{xcolor}
\usepackage{hyperref}

\usepackage{geometry}
\usepackage{algorithm}

\AtBeginDocument{%
  \providecommand\BibTeX{{%
    \normalfont B\kern-0.5em{\scshape i\kern-0.25em b}\kern-0.8em\TeX}}}

\copyrightyear{2025} 
\acmYear{2025} 
\setcopyright{acmlicensed}
\acmConference[SIGIR '25]{Proceedings of the 48th International ACM SIGIR Conference on Research and Development in Information Retrieval}{July 13-18, 2025}{Padua, Italy}
\acmBooktitle{Proceedings of the 48th International ACM SIGIR Conference on Research and Development in Information Retrieval (SIGIR '25), July 13-18, 2025, Padua, Italy}
\acmPrice{15.00}
\acmDOI{10.1145/3539618.3591976}
\acmISBN{978-1-4503-9408-6/23/07}

\begin{document}

\title{Towards Principled Learning for Re-ranking in Recommender Systems}

\author{Qunwei Li$^\ast$,\quad Linghui Li,\quad Jianbin Lin,\quad Wenliang Zhong}
\affiliation{Ant Group \country{China}  \\
\texttt{\{qunwei.qw,  hummy.llh, jianbin.ljb, yice.zwl,\}@utah.edu}
\vspace{1.5ex}
}
  \thanks{$^\ast$Work was done when Qunwei Li was at Ant Group. Now he is at Hechun Medical Technology Co., and can be connected by \texttt{lee880716@gmail.com.}}

\begin{abstract} 
\input{sections/abs}
\end{abstract}





\maketitle

\section{Introduction}
\input{sections/intro}

\section{Related Works}
\input{sections/related_works}
\section{Methodology}
\input{sections/methodology}

\section{Experiments}
\input{sections/exp}

\section{Conclusion}
\input{sections/conclusion}

\newpage
\bibliographystyle{ACM-Reference-Format}
\balance
\bibliography{main-sigir2023}

\end{document}

%% file: sections/abs.tex
As the final stage of recommender systems, re-ranking presents ordered item lists to users that best match their interests. It plays such a critical role and has become a trending research topic with much attention from both academia and industry. Recent advances of re-ranking are focused on attentive listwise modeling of interactions and mutual influences among items to be re-ranked. However, principles to guide the learning process of a re-ranker, and to measure the quality of the output of the re-ranker, have been always missing. In this paper, we study such principles to learn a good re-ranker. Two principles are proposed, including convergence consistency and adversarial consistency. These two principles can be applied in the learning of a generic re-ranker and improve its performance. We validate such a finding by various baseline methods over different datasets.

%% file: sections/intro.tex
Unlocking user insights and delivering personalized experiences in a huge variaty of Web applications like e-commence, social networks, and news feeds, Recommender Systems (RS) are widely adopted by many online service providers \cite{li2023edge,li2022prototypical,xie2022denoising,li2021learning}. In practice, the numbers of users and recommendable items can easily run into more than millions, especially for large online platforms \cite{eksombatchai2018pixie}. Detouring from directly recommending items to users using one model in a single stage, a realistic RS typically consists of several stages and models to serve user requests \cite{schafer1999recommender,liu2022neural}. The matching stage recalls or generates a smaller item pool comparing to all the recommendable items. Then, a ranking model scores the items from such a pool and first a few top-scored items are ranked and pulled into the stage of re-ranking. A re-ranker adjusts the final ranking result, which then would be presented to users. 

An item itself, and the ordering and mutual effect of the items in the same list, would both influence whether a user is interested in such an item. Thus, re-ranking aims at encoding a list of items into contextualized sequence representations and jointly ranking all the items at a time. Toward such understandings, recurrent neural networks (RNNs) are employ in GlobalRank \cite{zhuang2018globally} and DLCM \cite{ai2018learning} to encode the initial ranking list sequentially or bidirectionally. Comparing to RNNs, Transformer architecture \cite{vaswani2017attention} is more effective and efficient in modeling interactions of two items in an ordered sequence. Later studies started to encode the item list using Transformer. For example, PRM \cite{pei2019personalized} and SetRank \cite{pang2020setrank}, apply the self-attention mechanism in Transformers to modeling inter-dependencies among items. PEAR models item contexts from both the initial ranking list and the historical clicked item list \cite{li2022pear}.  RAISE \cite{lin2022attention} proposes to model individual attention weights to improve personalization. Please also refer to the comprehensive review in \cite{liu2022neural} and the references therein for recent developments in re-ranking.

In spite of the advances in the aforementioned explorations in modeling of a re-ranker, an important question regarding the learning of a re-ranker has never been answered: how can we tell the results yielded by the re-ranker is good enough in the learning process? Till now, the only answer one can provide is probably the accuracy metrics, e.g., NDCG (Normalized Discounted Cumulative Gain) or MAP (Mean Average Precision) of the re-ranking list. {Such metrics measure the goodness of fitting of a re-ranker over the test dataset after training, and can not be directly used to guide the learning process during training itself. }

We study and propose principles to answer the above question. Notice that listwise modeling of a re-ranker is quite different from conventional pointwise ranking, as it has input and output both in the form of an ordered list. We investigate the difference between the input list and output list, and propose two principles for a good re-ranker as:  \p1: \textbf{\emph{The resultant re-ranked list cannot be further re-ranked by the re-ranker}}; 
 \p2: \textbf{\emph{A small perturbation in the initial order of the ranking list cannot alter the result of the re-ranking list.}} 
 We term the principles as of \textbf{Convergence Consistency} and \textbf{Adversarial Consistency}.
 
We provide an algorithm to apply these two principles in the training of a re-ranker. No extra modeling or computation is needed, and the principles can be integrated into the loss for training a general re-ranker and improve its performance.

%% file: sections/related_works.tex
Among the earliest attempts of neural re-ranking methods, DLCM \cite{ai2018learning} MiDNN \cite{zhuang2018globally} propose to employ recurrent neural networks to encode the ranking list. Specifically, DLCM employs gated recurrent units (GRU) to sequentially encode the ranking results using their feature vectors, learning a local context model and use it to re-rank the ranking list. MiDNN also applies recurrent networks, the long-short term memory (LSTM), with a global feature extension method to incorporate mutual influences into the features of an item. 
 It formulates the re-ranking as a sequence generation problem, and sequentially selects the next items to form a best list possible.
PRM \cite{pei2019personalized} is one of the first pioneers to encode the ranking list with Transformer. By a straightforward adoption of the self-attention structure, which is a stack of multiple blocks of self-attention layers and feed-forward networks, the mutual influence between item pairs in the ranking list can be better explored than that by RNNs.
PEAR \cite{li2022pear} makes several major improvements over
the existing methods. Specifically, PEAR not only captures feature-level and item-level interactions, but also models item contexts
from both the initial ranking list and the historical clicked item list.
In addition to item-level ranking score prediction, it also augments
the training of PEAR with a list-level classification task to assess
users’ satisfaction on the whole ranking list.
To further improve personalization in re-ranking, a more recent work RAISE \cite{lin2022attention} maintains individual attention weights in modeling cross-item interactions for each user.

A group-wise scoring function is proposed in GSF \cite{ai2019learning}, which is then devised with DNN on all the size-$m$ permutations of items in the initial list of length $n$ ($m\le n$) for training. SetRank \cite{pang2020setrank} applies a variant of self-attention structure without positional encoding to convert an order list to a set,  preserving the permutation invariant property. GSF explicitly forms all possible permutations and SetRank deprives ordering from the initial ranking list, both learning permutation-invariant re-ranking.

%% file: sections/methodology.tex
Following recent works of re-ranking like \cite{li2022pear,pei2019personalized}, we first show in this section the general modeling of a re-ranker. Then, we reason and provide principles to guide the learning process of a re-ranker, and to measure the quality of the output of the re-ranker. An algorithm is finally presented to integrate the proposed principles.
\subsection{General Model of Re-ranker}
The architecture of a general re-ranking model 
consists of four major parts: the Input Layer, the Feature Interaction Modeling, the Item interaction Modeling,and the
Output Layer. The model takes in an initial ordered list of items generated by a
ranking model and features of a user as input and yields scores of each item in the list for re-ranking. The detailed structure will be introduced separately as follows.

\textbf{Input Layer.} 
The goal of the input layer is to prepare comprehensive representations of all items in the initial list and feed it to the encoding layer. First we have a fixed length of initial sequential list $\s = [i_1, i_2, \ldots, i_n]$
 given by the previous ranking method. Same
as the previous ranking method, we have a raw feature matrix $\X\in \R ^{n\times d_{item}}$. Each row in $\X$ represents the raw feature vector $\x_i$
for an item $i\in\s$. One may encode item feature vector and apply learnable vector over $\x_i$ for better personalization \cite{pei2019personalized}. Without loss of generality, we uniformly use $\X$ to denote item-specific features. Some re-ranking works also consider the modeling of historical user-item interactions and other user-specific features \cite{li2022pear}. We denote such inputs as user feature and represent it by $\bm{U}$.

\textbf{Feature Interaction Modeling.} In order to utilize the sequential
information in the initial list, a position $\pe \in \R^{n}$ is injected
into the input embedding. Then the item feature
matrix is $\X+\pe$. Note that one does not necessarily need position embedding if she uses RNNs to encode mutual influence of the order of the initial re-ranking list, while the list is sequentially input to RNN units and position $\pe$ is still used.

Apart from position embedding, other cross-item feature manipulations could also be applied for better extracting useful information here, and we omit special articulation for clearer presentation as it is not the focus of this paper.

\textbf{Item Interaction Modeling.}
The goal of the item interaction modeling layer is to encode the mutual influences of item-pairs and other item-specific information, and the  order of the initial ranking list. To achieve such a goal, one may adopt an RNN or Transformer based encoder for it should have the ability to effectively process sequential data, as the ordering in the initial ranking list contains crucial information. The self-attention mechanism in Transformer is particularly preferable 
in the re-ranking tasks as it can effectively model the mutual influences
for any two items so as to capture the influence of presenting one item to a user on liking any other items later presented. 

\textbf{Output Layer.}
After item interaction modeling, and processing of user feature, all the  embeddings would be concatenated and fed into simple  multi-layer perceptron (MLP) layers followed by a softmax layer.
The objective of the output layer is to generate a re-ranking score for
each item $i = i_1, \ldots ,i_n$. The final output is the probability of click/like for each item, which is expressed as
$P(y_i|\X,\U ;\theta)$, where $y_i \in \y$ is the label of click-through for item $i$ and the whole network is parameterized by $\theta$. 

Typically, a log-loss is used to train the network and is shown as
\begin{align}
\loss(\X,\pe) = -\sum_{u}\sum_{i\in \s_u} y_i\log(P(yi
|\X, \U ;\theta) ),
\end{align}
which is summed over user $u$ for all the lists $\s_u$ serving user requests. Then, one uses $P(yi|\X, \U ;\theta)$ as the score to re-rank the items. 

\subsection{Proposed Principles}
Starting from here, we propose two principles of \textbf{Convergence Consistency} and \textbf{Adversarial Consistency} that can  guide the learning process of a re-ranker, and also measure the quality of the output of the re-ranker. We first provide corresponding reasoning and rationale, and then give an algorithm to implement the principles in the training of a general re-ranker.

\textbf{Reasoning and Rationale.}
\p1: \textbf{\emph{The resultant re-ranked list cannot be further re-ranked by the re-ranker.}} The essence of a re-ranker is based on the following simple logic flow: one (RS) asks a re-ranker whether a list $A$ is well ordered to be presented to a user, and the re-ranker would come up with an ordered resultant list $B$. If one then asks the re-ranker whether $B$ is well ordered, the re-ranker comes up with an ordered resultant list $C$. In such a flow of $A\xrightarrow{\text{re-ranker}} B \xrightarrow{\text{re-ranker}} C$, if $B\neq C$, then we declare that the resultant list can be further re-ranked by the re-ranker and such a re-ranker is not trained well as it is not even confident and decided with its own output. On the contrary, if a re-ranker follows a flow of $A\xrightarrow{\text{re-ranker}} B \xrightarrow{\text{re-ranker}} B$, one can trust the re-ranker and present the re-ranked result to the user. 

 \p2: \textbf{\emph{A small perturbation in the initial order of the ranking list cannot alter the result of the re-ranking list.}} The initial list fed to a re-ranker is typically ordered by a ranker, which is trained with user-item interactions. The order in the list before re-ranking has crucial information provided by the ranker, and position of items in such an initial list is taken into consideration in modeling a re-ranker in practice. Works like \cite{pang2020setrank} completely ignore such information cannot fully explore the item interactions. A good re-ranker takes in two lists $A$ and $B$ with same items and different order, should output a same list $C$, as $\{A,B\} \xrightarrow{\text{re-ranker}} C$, which is the foundation of \p2. On top of such a logic, notice that the order provided by a ranker in the initial list is crucial, while it also has noise since it is nevertheless not the ground truth. In such a case, one injects a small perturbation in the input list, and the re-ranker should yield the same result, bearing consistency and robustness that mimics the success of adversarial training \cite{bai2021recent}. In the experiments, such a small perturbation is realized by switching positions of two adjacent items. The principle emphasizes one and true list after reranking despite the initial order of the list.

\textbf{Algorithm.} We now present the algorithm to integrate the proposed principles into the learning of a general re-ranker. The re-ranker is denoted by $\rr(\X,\pe)$, where the re-ranker is parameterized by learnable neural networks and has inputs of feature $\X$ and the initial position of the items $\pe$. The output of the re-ranker is a vector of scores $\bm p$ representing a new ordering position ${\pe}^\prime$ of the initial item list. We use the following formula to express the above process as:
\begin{align}\label{eq:p1}
    {\pe}^\prime \sim \bm p =P(\y
|\X, \pe ;
\theta) = \rr(\X,\pe)
\end{align}
Recall principle $\p1$ and the position of a further re-ranked list can be expressed as
\begin{align}\label{eq:p2}
    {\pe}^{\prime\prime} \sim  \rr(\X,{\pe}^\prime),
\end{align}
which then should also be reflected in the loss to fit the labels and the final loss now is
\begin{align}
    \loss = \loss(\X,\pe)+\loss(\X,\pe^\prime).
\end{align}

Again, recalling principle $\p2$, we randomly select two adjacent items in the initial list  and switch their positions so as to add a small perturbation and obtain a new list position as $\hat{\pe}$.  Similarly, the re-ranker with such a position input should also fits the label well and the loss goes
\begin{align}
    \loss = \loss(\X,\pe)+\loss(\X,\pe^\prime)+\loss(\X,\hat{\pe}).
\end{align}

Following the essence of the principles, the outputs of the re-ranker with different positions should remain the same. We propose here a new loss to promote such similarity, and name it by Contrastive Similarity (CS). Let the position represented as a vector of indexed numbers $[0,1,\ldots,n-1]$, and item $i$ is $\pe_i$-th in the list. The CS loss of two different orderings of two lists $\pe_A$ and $\pe_B$ is
\begin{align}
    \loss_{CS}(\pe_A, \pe_B) = |\pe_A- \pe_B|^T (\rr(\X,\pe_A)-\rr(\X,\pe_B))^2.
\end{align}
Here the square operation is element wise. Such a loss is physically a weighted square error of two re-ranking scores, and the weight is the difference in two positions resulting from two re-ranking calls. It only penalizes the case where the ordering of two lists differs.

The principles proposed concretely regularizes the ordering difference of two lists and we add CS loss to training as
\begin{align}\label{loss}
    \loss = \loss(\X,\pe)+\underbrace{\loss(\X,\pe^\prime)+\loss_{CS}(\pe^{\prime\prime}, \pe^\prime)}_\text{\p1}+\underbrace{\loss(\X,\hat{\pe})+\loss_{CS}(\hat \pe, \pe^\prime)}_\text{\p2}.
\end{align}
\begin{algorithm}[t]
    \caption{Algorithm of Principled Learning for Re-ranking}\label{algorithm}
    \begin{algorithmic}[1]
        \STATE Input samples $\X,\pe$ 
        \STATE Instantiate the model of a re-ranker as $\rr()$
        \FOR{epochs = $1$ to $N$}
                \STATE Randomly select two adjacent items and switch their positions in $\pe$, and obtain $\hat\pe$
                \STATE Obtain position from initial input as in Eq. \ref{eq:p1}
                \STATE Obtain further re-ranked position as in Eq. \ref{eq:p2}
                \STATE Update the re-ranker parameterized by $\theta$ with loss $\loss$ as in Eq. \ref{loss}
        \ENDFOR
    \end{algorithmic}
\end{algorithm}
We provide in Algorithm \ref{algorithm} the whole training process for easier understanding. Note that proposed principles also apply to re-ranking with an evaluator-generator paradigm \cite{wang2019sequential} since the evaluator perfectly follows the re-ranking framework that we study.

\begin{table*}[tb]
\centering
\resizebox{\linewidth}{!}{
\begin{tabular}{llllllllllll}
\cline{1-11}
        & AUC            & NDCG           & MAP@5            & MAP@10           & MAP@15 & MAP@20 & Precision@5            & Precision@10 & Precision@15 & Precision@20 &  \\ \hline
DLCM[\citeyear{ai2018learning}]    & 0.5805+1.23\%                         & 0.4771+0.13\%                         & 0.3044+0.16\%                         & 0.3141+0.15\%                         & 0.3102+0.15\%                         & 0.3045+0.17\%                         & 0.1340+0.33\% & 0.1050+0.43\% & 0.0887+0.42\% & 0.0768+1.47\% \\
MiDNN[\citeyear{zhuang2018globally}]   & 0.6172+0.49\%                         & 0.4735+0.45\%                         & 0.2991+0.96\%                         & 0.3094+0.86\%                         & 0.3058+0.81\%                         & 0.3001+0.81\%                         & 0.1322+0.90\% & 0.0964+8.70\% & 0.0881+0.56\% & 0.0774+0.37\% \\
PRM[\citeyear{pei2019personalized}]     & 0.6161+0.41\% & 0.4750+0.29\% & 0.3012+0.64\% & 0.3113+0.57\% & 0.3076+0.55\% & 0.3019+0.54\% & 0.1331+0.55\% & 0.1047+0.41\% & 0.0885+0.34\% & 0.0776+0.26\% \\
SetRank[\citeyear{pang2020setrank}] & 0.6101+1.28\%                         & 0.4670+0.41\%                         & 0.2898+0.70\%                         & 0.3012+0.36\%                         & 0.2982+0.59\%                         & 0.2928+0.30\%                         & 0.1295+0.27\% & 0.1027+0.01\% & 0.0872+0.36\% & 0.0767+0.22\% \\
PEAR[2022]    & 0.6082+2.93\%                         & 0.4606+2.91\%                         & 0.2810+6.98\%                         & 0.2929+6.05\%                         & 0.2901+5.73\%                         & 0.2848+5.66\%                         & 0.1245+6.08\%  & 0.0996+4.13\% & 0.0857+2.54\% & 0.0762+1.30\% \\
RAISE[\citeyear{lin2022attention}]   & 0.6152+0.16\%                         & 0.4741+0.05\%                         & 0.2997+0.10\%                         & 0.3101+0.09\%                         & 0.3064+0.08\%                         & 0.3008+0.07\%                         & 0.1329+0.12\% & 0.1046+0.10\% & 0.0884+0.13\% & 0.0775+0.11\%\\\hline

\end{tabular}
}
\caption{Performance comparison of various methods with dataset \textbf{PRM Public}.}
\label{tab:prm}
\end{table*}



%% file: sections/exp.tex
In the experiments, we evaluate the performance comparison in terms of popular ranking metrics AUC, NDCG, MAP@$K$, and Precision@$K$  with $k=5,10,15,20$. The learning rate is tuned among $\{e^{-4},5e^{-5},e^{-5}\}$ with the optimizer of Adam to yield the best performance.
 We release the codes \href{https://github.com/goldenxingxing/principled_reranking}{\textit{here}}.
\subsection{Experimental Settings} \label{subsect:exp_setting}
We use the same datasets and feature processing methods as presented in \href{https://github.com/LibRerank-Community/LibRerank}{LibRerank}  library \cite{liu2022neural}.

\noindent\textbf{Datasets.} We conduct experiments on two public
recommendation datasets, \textbf{Ad} and \textbf{PRM Public}.
The original Ad dataset records 1 million users and 26 million ad display/click logs, with 8 user profiles, 6 item features. The records of each user are formed into ranking lists according to the timestamp of the user browsing the advertisement that are within five minutes.
The original \textbf{PRM Public} dataset contains re-ranking lists from a real-world e-commerce RS. Each record is a recommendation list consisting of 3 user profile features, 5 categorical, and 19 dense item features. 

\noindent\textbf{Baselines.} We experiment with the methods mentioned in related work from Section 2. 

The initial ranking list is provided by LambdaMart \cite{burges2010ranknet}, with a length of 10 and 30, and the number of training epoch is 100 and 30 for the two respective datasets. For the dataset of \textbf{AD}, we measure the performance by AUC, NDCG and MAP@k, and we provide extra Precision@k for the more complex dataset of \textbf{PRM Public}.

\subsection{Quantitative Comparison}
\begin{table}[h]
\centering
\resizebox{\columnwidth}{!}{
\begin{tabular}{lllll}
\hline
        & AUC           & NDCG          & MAP@5           & MAP@10          \\ \hline
DLCM  & 0.8149+0.58\% & 0.6923+0.60\% & 0.5962+0.82\% & 0.5987+1.05\% \\
MiDNN   & 0.8444+0.74\% & 0.6943+0.61\% & 0.5980+0.96\% & 0.6018+0.96\% \\
PRM     & 0.8461+0.36\% & 0.6928+1\%    & 0.5959+1.57\% & 0.6000+1.54\% \\
SetRank & 0.8181+1.93\% & 0.6940+0.79\% & 0.5972+1.27\% & 0.6010+1.29\% \\
PEAR   & 0.8133+1.93\% & 0.6860+1.03\% & 0.5794+2.78\% & 0.5837+2.79\% \\
RAISE   & 0.8163+3.88\% & 0.6886+0.65\% & 0.5841+2.12\% & 0.5882+2.06\%\\\hline
\end{tabular}}
\caption{Performance comparison of various methods with dataset \textbf{AD}.}
\label{tab:ad}
\end{table}
We report the mean of the metric from 10 trials and omit the standard deviation (STD) due to the space limit and the fact that we find STD is mostly in small orders of $e^{-3}\sim e^{-5}$. The results are shown in the form of Metric $X$ by baseline plus Improvement $Y\%$, which reads the baseline method has a performance metric value of $X$ and is then improved by $Y\%$ when the proposed principles are integrated in the learning of the method using Algorithm~\ref{algorithm}. We can see from Table \ref{tab:prm} and Table \ref{tab:ad} that all the baseline methods with the proposed principles achieve performance improvement, ranging from 0.05\% to 9.89\%, showing the effectiveness and feasibility of the principles on top of various methods and different metric measures. We see no obvious evidence of the improvements being correlated with certain methods or metrics, showing the universal robustness of the principles over methods and metrics.

\subsection{Ablation Study}
We present here the ablation performance with the dataset \textbf{AD}. We provide here two aspects into the ablation study.
\subsubsection{Performance Imporvement}
Since we have obtained performance improvement with both the principles $\p1$ and $\p2$ integrated, we in Table~\ref{tab:ablation} show the effect on the improvement of implementing either one of the principles. The form of Method-Principle in the table means the case where the Method is integrated with the Principle. The results are calculated as ratio of performance improvement with implementing one principle over that with two principles. We can see from the table that either one of the two principles can help to improve the performance of the re-ranking methods. Most of the individual improvement by one principle is less than 100\%, meaning modeling with both the principles would yield better results. Some of the individual improvement is already larger than 100\%, and if we sum up the two improvements for one certain baseline, we find that the result is irrelevant of 100\% as well, indicating that the two principles shed light into orthogonal design for performance improvement. 
\begin{table}[h]
\centering
\resizebox{\columnwidth}{!}{
\begin{tabular}{lllll}
\hline
           & AUC       & NDCG       & M@5        & M@10       \\ \hline
DLCM-$\p1$    & 84.0099\% & 41.4273\%  & 45.7731\%  & 56.1592\%  \\
DLCM-$\p2$    & 27.2812\% & 37.9339\%  & 40.1911\%  & 52.1466\%  \\ \hline
MiDNN-$\p1$   & 55.4928\% & 0.7853\%   & 1.1346\%   & 2.9275\%   \\
MiDNN-$\p2$   & 67.4724\% & 20.0534\%  & 23.8126\%  & 25.0348\%  \\ \hline
PRM-$\p1$     & 42.1215\% & 9.8880\%   & 131.0388\% & 9.5619\%   \\
PRM-$\p2$     & 44.4743\% & 5.9221\%   & 10.9997\%  & 8.2492\%   \\ \hline
SetRank-$\p1$ & 11.9313\% & 26.0249\%  & 36.2582\%  & 58.9121\%  \\
SetRank-$\p2$ & 33.1463\% & 37.0095\%  & 37.1925\%  & 46.1428\%  \\ \hline
PEAR-$\p1$    & 36.3606\% & 6.7894\%   & 31.0654\%  & 29.7227\%  \\
PEAR-$\p2$    & 40.4980\% & 32.5201\%  & 44.4642\%  & 43.9986\%  \\ \hline
RAISE-$\p1$   & 3.3862\%  & 44.7252\%  & 74.7007\%  & 75.0186\%  \\
RAISE-$\p2$   & 5.8047\%  & 74.4366\%  & 85.7724\%  & 87.0452\%  \\ \hline
\end{tabular}}
\caption{Principle ablation performance of various methods.}
\label{tab:ablation}
\end{table}
\begin{table}[]
\begin{tabular}{lllll}\hline
        & \multicolumn{2}{l}{ $\p1$ Obedience} & \multicolumn{2}{l}{$\p2$ Obedience}       \\ \hline
        & baseline      & baseline+$\p1$      & baseline & baseline+$\p2$                 \\ \hline
DLCM    & 0.5708        & 0.8166              & 0.6235   & 0.7769                         \\
MiDNN   & 0.3015        & 0.3938              & 0.3059   & 0.4168 \\
PRM     & 0.4074        & 0.4576              & 0.3803   & 0.3992                         \\
SetRank & 0.5106        & 0.5665              & 0.4413   & 0.4861                         \\
PEAR    & 0.032         & 0.0450              & 0.0303   & 0.0440                         \\
RAISE   & 0.3988        & 0.4295              & 0.3093   & 0.4271   \\ \hline                     
\end{tabular}
\caption{Principle obedience of various methods.}
\label{tab:obedience}
\end{table}
\subsubsection{Principle Obedience}
With performance improvement at hand by the proposed method, a straightforward concern needs resolving is whether the improvement sits alongside an increased scale of principle obedience. We report the percentage of samples in test set of \textbf{AD} that obey the proposed principles. Essentially, for the flow of $A\xrightarrow{\text{re-ranker}} B \xrightarrow{\text{re-ranker}} C$ in $\p1$, $\p1$ is obeyed only if $B=C$, and similar goes for $\p2$. It is shown in Table~\ref{tab:obedience} that the baseline methods learned with the respective principle integrated yield increased obedience to the corresponding principle. 

%% file: sections/conclusion.tex
In this paper, two principles to guide the learning process of a re-ranker, and to measure the quality of the output of the re-ranker have been proposed, termed as convergence consistency and adversarial consistency respectively. We provided the reasoning and rationale for the two principles and an algorithm to integrate the principles into the training of a general re-ranker to regularize the output of the re-ranker so as to better model the mutual influences between item pairs in the ranking list. Such principles can be off-the-shelf add-ons to a re-ranker and improve its performance. To the best of our knowledge, this is the first work towards principled learning of a re-ranker.

%% file: main.bbl

\begin{thebibliography}{18}


\ifx \showCODEN    \undefined \def \showCODEN     #1{\unskip}     \fi
\ifx \showDOI      \undefined \def \showDOI       #1{#1}\fi
\ifx \showISBNx    \undefined \def \showISBNx     #1{\unskip}     \fi
\ifx \showISBNxiii \undefined \def \showISBNxiii  #1{\unskip}     \fi
\ifx \showISSN     \undefined \def \showISSN      #1{\unskip}     \fi
\ifx \showLCCN     \undefined \def \showLCCN      #1{\unskip}     \fi
\ifx \shownote     \undefined \def \shownote      #1{#1}          \fi
\ifx \showarticletitle \undefined \def \showarticletitle #1{#1}   \fi
\ifx \showURL      \undefined \def \showURL       {\relax}        \fi
\providecommand\bibfield[2]{#2}
\providecommand\bibinfo[2]{#2}
\providecommand\natexlab[1]{#1}
\providecommand\showeprint[2][]{arXiv:#2}

\bibitem[Ai et~al\mbox{.}(2018)]%
        {ai2018learning}
\bibfield{author}{\bibinfo{person}{Qingyao Ai}, \bibinfo{person}{Keping Bi}, \bibinfo{person}{Jiafeng Guo}, {and} \bibinfo{person}{W~Bruce Croft}.} \bibinfo{year}{2018}\natexlab{}.
\newblock \showarticletitle{Learning a deep listwise context model for ranking refinement}. In \bibinfo{booktitle}{\emph{The 41st international ACM SIGIR conference on research \& development in information retrieval}}. \bibinfo{pages}{135--144}.
\newblock


\bibitem[Ai et~al\mbox{.}(2019)]%
        {ai2019learning}
\bibfield{author}{\bibinfo{person}{Qingyao Ai}, \bibinfo{person}{Xuanhui Wang}, \bibinfo{person}{Sebastian Bruch}, \bibinfo{person}{Nadav Golbandi}, \bibinfo{person}{Michael Bendersky}, {and} \bibinfo{person}{Marc Najork}.} \bibinfo{year}{2019}\natexlab{}.
\newblock \showarticletitle{Learning groupwise multivariate scoring functions using deep neural networks}. In \bibinfo{booktitle}{\emph{Proceedings of the 2019 ACM SIGIR international conference on theory of information retrieval}}. \bibinfo{pages}{85--92}.
\newblock


\bibitem[Bai et~al\mbox{.}(2021)]%
        {bai2021recent}
\bibfield{author}{\bibinfo{person}{Tao Bai}, \bibinfo{person}{Jinqi Luo}, \bibinfo{person}{Jun Zhao}, \bibinfo{person}{Bihan Wen}, {and} \bibinfo{person}{Qian Wang}.} \bibinfo{year}{2021}\natexlab{}.
\newblock \showarticletitle{Recent advances in adversarial training for adversarial robustness}.
\newblock \bibinfo{journal}{\emph{arXiv preprint arXiv:2102.01356}} (\bibinfo{year}{2021}).
\newblock


\bibitem[Burges(2010)]%
        {burges2010ranknet}
\bibfield{author}{\bibinfo{person}{Christopher~JC Burges}.} \bibinfo{year}{2010}\natexlab{}.
\newblock \showarticletitle{From ranknet to lambdarank to lambdamart: An overview}.
\newblock \bibinfo{journal}{\emph{Learning}} \bibinfo{volume}{11}, \bibinfo{number}{23-581} (\bibinfo{year}{2010}), \bibinfo{pages}{81}.
\newblock


\bibitem[Eksombatchai et~al\mbox{.}(2018)]%
        {eksombatchai2018pixie}
\bibfield{author}{\bibinfo{person}{Chantat Eksombatchai}, \bibinfo{person}{Pranav Jindal}, \bibinfo{person}{Jerry~Zitao Liu}, \bibinfo{person}{Yuchen Liu}, \bibinfo{person}{Rahul Sharma}, \bibinfo{person}{Charles Sugnet}, \bibinfo{person}{Mark Ulrich}, {and} \bibinfo{person}{Jure Leskovec}.} \bibinfo{year}{2018}\natexlab{}.
\newblock \showarticletitle{Pixie: A system for recommending 3+ billion items to 200+ million users in real-time}. In \bibinfo{booktitle}{\emph{Proceedings of the 2018 world wide web conference}}. \bibinfo{pages}{1775--1784}.
\newblock


\bibitem[Li et~al\mbox{.}(2022a)]%
        {li2022prototypical}
\bibfield{author}{\bibinfo{person}{Ningning Li}, \bibinfo{person}{Qunwei Li}, \bibinfo{person}{Xichen Ding}, \bibinfo{person}{Shaohu Chen}, {and} \bibinfo{person}{Wenliang Zhong}.} \bibinfo{year}{2022}\natexlab{a}.
\newblock \showarticletitle{Prototypical Contrastive Learning and Adaptive Interest Selection for Candidate Generation in Recommendations}. In \bibinfo{booktitle}{\emph{Proceedings of the 31st ACM International Conference on Information \& Knowledge Management}}. \bibinfo{pages}{4183--4187}.
\newblock


\bibitem[Li et~al\mbox{.}(2021)]%
        {li2021learning}
\bibfield{author}{\bibinfo{person}{Youru Li}, \bibinfo{person}{Xiaobo Guo}, \bibinfo{person}{Wenfang Lin}, \bibinfo{person}{Mingjie Zhong}, \bibinfo{person}{Qunwei Li}, \bibinfo{person}{Zhongyi Liu}, \bibinfo{person}{Wenliang Zhong}, {and} \bibinfo{person}{Zhenfeng Zhu}.} \bibinfo{year}{2021}\natexlab{}.
\newblock \showarticletitle{Learning dynamic user interest sequence in knowledge graphs for click-through rate prediction}.
\newblock \bibinfo{journal}{\emph{IEEE Transactions on Knowledge and Data Engineering}} \bibinfo{volume}{35}, \bibinfo{number}{1} (\bibinfo{year}{2021}), \bibinfo{pages}{647--657}.
\newblock


\bibitem[Li et~al\mbox{.}(2022b)]%
        {li2022pear}
\bibfield{author}{\bibinfo{person}{Yi Li}, \bibinfo{person}{Jieming Zhu}, \bibinfo{person}{Weiwen Liu}, \bibinfo{person}{Liangcai Su}, \bibinfo{person}{Guohao Cai}, \bibinfo{person}{Qi Zhang}, \bibinfo{person}{Ruiming Tang}, \bibinfo{person}{Xi Xiao}, {and} \bibinfo{person}{Xiuqiang He}.} \bibinfo{year}{2022}\natexlab{b}.
\newblock \showarticletitle{PEAR: Personalized Re-ranking with Contextualized Transformer for Recommendation}. In \bibinfo{booktitle}{\emph{Companion Proceedings of the Web Conference 2022}}. \bibinfo{pages}{62--66}.
\newblock


\bibitem[Li et~al\mbox{.}(2023)]%
        {li2023edge}
\bibfield{author}{\bibinfo{person}{Zexi Li}, \bibinfo{person}{Qunwei Li}, \bibinfo{person}{Yi Zhou}, \bibinfo{person}{Wenliang Zhong}, \bibinfo{person}{Guannan Zhang}, {and} \bibinfo{person}{Chao Wu}.} \bibinfo{year}{2023}\natexlab{}.
\newblock \showarticletitle{Edge-cloud Collaborative Learning with Federated and Centralized Features}.
\newblock \bibinfo{journal}{\emph{SIGIR}} (\bibinfo{year}{2023}).
\newblock


\bibitem[Lin et~al\mbox{.}(2022)]%
        {lin2022attention}
\bibfield{author}{\bibinfo{person}{Zhuoyi Lin}, \bibinfo{person}{Sheng Zang}, \bibinfo{person}{Rundong Wang}, \bibinfo{person}{Zhu Sun}, \bibinfo{person}{J Senthilnath}, \bibinfo{person}{Chi Xu}, {and} \bibinfo{person}{Chee~Keong Kwoh}.} \bibinfo{year}{2022}\natexlab{}.
\newblock \showarticletitle{Attention over self-attention: Intention-aware re-ranking with dynamic transformer encoders for recommendation}.
\newblock \bibinfo{journal}{\emph{IEEE Transactions on Knowledge and Data Engineering}} (\bibinfo{year}{2022}).
\newblock


\bibitem[Liu et~al\mbox{.}(2022)]%
        {liu2022neural}
\bibfield{author}{\bibinfo{person}{Weiwen Liu}, \bibinfo{person}{Yunjia Xi}, \bibinfo{person}{Jiarui Qin}, \bibinfo{person}{Fei Sun}, \bibinfo{person}{Bo Chen}, \bibinfo{person}{Weinan Zhang}, \bibinfo{person}{Rui Zhang}, {and} \bibinfo{person}{Ruiming Tang}.} \bibinfo{year}{2022}\natexlab{}.
\newblock \showarticletitle{Neural re-ranking in multi-stage recommender systems: A review}.
\newblock \bibinfo{journal}{\emph{IJCAI}} (\bibinfo{year}{2022}).
\newblock


\bibitem[Pang et~al\mbox{.}(2020)]%
        {pang2020setrank}
\bibfield{author}{\bibinfo{person}{Liang Pang}, \bibinfo{person}{Jun Xu}, \bibinfo{person}{Qingyao Ai}, \bibinfo{person}{Yanyan Lan}, \bibinfo{person}{Xueqi Cheng}, {and} \bibinfo{person}{Jirong Wen}.} \bibinfo{year}{2020}\natexlab{}.
\newblock \showarticletitle{Setrank: Learning a permutation-invariant ranking model for information retrieval}. In \bibinfo{booktitle}{\emph{Proceedings of the 43rd international ACM SIGIR conference on research and development in information retrieval}}. \bibinfo{pages}{499--508}.
\newblock


\bibitem[Pei et~al\mbox{.}(2019)]%
        {pei2019personalized}
\bibfield{author}{\bibinfo{person}{Changhua Pei}, \bibinfo{person}{Yi Zhang}, \bibinfo{person}{Yongfeng Zhang}, \bibinfo{person}{Fei Sun}, \bibinfo{person}{Xiao Lin}, \bibinfo{person}{Hanxiao Sun}, \bibinfo{person}{Jian Wu}, \bibinfo{person}{Peng Jiang}, \bibinfo{person}{Junfeng Ge}, \bibinfo{person}{Wenwu Ou}, {et~al\mbox{.}}} \bibinfo{year}{2019}\natexlab{}.
\newblock \showarticletitle{Personalized re-ranking for recommendation}. In \bibinfo{booktitle}{\emph{Proceedings of the 13th ACM conference on recommender systems}}. \bibinfo{pages}{3--11}.
\newblock


\bibitem[Schafer et~al\mbox{.}(1999)]%
        {schafer1999recommender}
\bibfield{author}{\bibinfo{person}{J~Ben Schafer}, \bibinfo{person}{Joseph Konstan}, {and} \bibinfo{person}{John Riedl}.} \bibinfo{year}{1999}\natexlab{}.
\newblock \showarticletitle{Recommender systems in e-commerce}. In \bibinfo{booktitle}{\emph{Proceedings of the 1st ACM conference on Electronic commerce}}. \bibinfo{pages}{158--166}.
\newblock


\bibitem[Vaswani et~al\mbox{.}(2017)]%
        {vaswani2017attention}
\bibfield{author}{\bibinfo{person}{Ashish Vaswani}, \bibinfo{person}{Noam Shazeer}, \bibinfo{person}{Niki Parmar}, \bibinfo{person}{Jakob Uszkoreit}, \bibinfo{person}{Llion Jones}, \bibinfo{person}{Aidan~N Gomez}, \bibinfo{person}{{\L}ukasz Kaiser}, {and} \bibinfo{person}{Illia Polosukhin}.} \bibinfo{year}{2017}\natexlab{}.
\newblock \showarticletitle{Attention is all you need}.
\newblock \bibinfo{journal}{\emph{Advances in neural information processing systems}}  \bibinfo{volume}{30} (\bibinfo{year}{2017}).
\newblock


\bibitem[Wang et~al\mbox{.}(2019)]%
        {wang2019sequential}
\bibfield{author}{\bibinfo{person}{Fan Wang}, \bibinfo{person}{Xiaomin Fang}, \bibinfo{person}{Lihang Liu}, \bibinfo{person}{Yaxue Chen}, \bibinfo{person}{Jiucheng Tao}, \bibinfo{person}{Zhiming Peng}, \bibinfo{person}{Cihang Jin}, {and} \bibinfo{person}{Hao Tian}.} \bibinfo{year}{2019}\natexlab{}.
\newblock \showarticletitle{Sequential evaluation and generation framework for combinatorial recommender system}.
\newblock \bibinfo{journal}{\emph{arXiv preprint arXiv:1902.00245}} (\bibinfo{year}{2019}).
\newblock


\bibitem[Xie et~al\mbox{.}(2022)]%
        {xie2022denoising}
\bibfield{author}{\bibinfo{person}{Sicong Xie}, \bibinfo{person}{Qunwei Li}, \bibinfo{person}{Weidi Xu}, \bibinfo{person}{Kaiming Shen}, \bibinfo{person}{Shaohu Chen}, {and} \bibinfo{person}{Wenliang Zhong}.} \bibinfo{year}{2022}\natexlab{}.
\newblock \showarticletitle{Denoising Time Cycle Modeling for Recommendation}. In \bibinfo{booktitle}{\emph{Proceedings of the 45th International ACM SIGIR Conference on Research and Development in Information Retrieval}}. \bibinfo{pages}{1950--1955}.
\newblock


\bibitem[Zhuang et~al\mbox{.}(2018)]%
        {zhuang2018globally}
\bibfield{author}{\bibinfo{person}{Tao Zhuang}, \bibinfo{person}{Wenwu Ou}, {and} \bibinfo{person}{Zhirong Wang}.} \bibinfo{year}{2018}\natexlab{}.
\newblock \showarticletitle{Globally optimized mutual influence aware ranking in e-commerce search}.
\newblock \bibinfo{journal}{\emph{IJCAI}} (\bibinfo{year}{2018}).
\newblock


\end{thebibliography}
